\documentclass[prb,
twocolumn,
superscriptaddress,showpacs,amsmath,amssymb]{revtex4}

\begin{document}

\title{Negative temperature: further extensions}

\author{G.E.~Volovik}
\affiliation{Low Temperature Laboratory, Aalto University,  P.O. Box 15100, FI-00076 Aalto, Finland}
\affiliation{Landau Institute for Theoretical Physics, acad. Semyonov av., 1a, 142432,
Chernogolovka, Russia}

\date{\today}

\begin{abstract}
This paper is motivated by the recent paper M. Baldovin, S. Iubini, R. Livi and A. Vulpiani, Statistical mechanics of systems with negative temperature, arXiv:2103.12572. The authors suggest that negative absolute temperatures are consistent with equilibrium thermodynamics. This is correct, if the environment also has negative temperature. Otherwise such states represent the transient though maybe long-lived metastable effects. Here we consider the extension of the negative temperatures to the relativistic systems. 
We show that the negative state is possible in the Dirac vacuum, which is obtained by the $PT$ symmetry transformation from the conventional Dirac sea. If such vacuum with inverse population fills the whole Universe, its thermodynamics determined by negative temperatures becomes equilibrium.
\end{abstract}
\pacs{
}

\maketitle

% \tableofcontents
 \newpage

%\twocolumn

\section{Introduction}

The authors of recent paper \cite{Baldovin2021} suggest that negative absolute temperatures are consistent with equilibrium thermodynamics. All the  thermodynamic properties, such as thermometry, thermodynamics of cyclic transformations, ensemble equivalence, fluctuation-dissipation relations, response theory and the transport processes, can be reformulated to include the
negative temperatures. In Ref. \cite{Baldovin2021} the condensed matter systems are discussed, such as the subsystem of nuclear spins with inverse spin population. The states with negative temperature has been experimentally studied in detail, and even the magnetic  phase transitions occurring at negative temperature have been detected.\cite{Lounasmaa1997}

The equilibrium thermodynamics at $T<0$ is only possible, if the environment also has negative temperature. Otherwise,
the heat will be transversed from the negative temperature system to the environment, and the whole system will relax to the 
conventional state with $T>0$. Here we show that the negative temperature states are also possible for the quantum vacuum in the relativistic quantum field theories. And if this vacuum state fills the whole Universe, this vacuum becomes thermodynamically stable in agreement with the conjecture in Ref. \cite{Baldovin2021}. 

\section{PT symmetry in relativistic quantum fields}

The relativistic vacuum with inverse population can be obtained by the $PT$ symmetry operation, where $P$  and $T$  are space and time reversal transformations correspondingly. When the gravitational tetrads are taken into account, the discrete  $PT$ symmetry of spacetime acquires new formulations, since the sign of the tetrad field becomes also important. \cite{Rovelli2012a,Boyle2018,Boyle2018b,Vergeles2021,Zubkov2021} In particular, if the tetrads are composite objects, which are formed as bilinear combinations of the fermionic operators,\cite{Diakonov2011,Diakonov2012,ObukhovHehl2012} the parity $P$ and the time reversal $T$ operations
are formed as combinations of the original more fundamental symmetry operators,  $P=P_{\rm c}  P_{\rm s}$ and  $T=T_{\rm c} T_{\rm s}$. Here  $P_{\rm c}$ and $T_{\rm c}$ refer to pure coordinate transformations, $P_{\rm c} {\bf r} = -  {\bf r}$ and $T_{\rm c} t = - t$, while $P_{\rm s}$ and $T_{\rm s}$ refer to the corresponding transformations of Dirac or Weyl spinors.\cite{Volovik2020b}
In this approach the Lorentz transformation also represents  the combination of the fundamental symmetry operations, $L=L_{\rm c}  L_{\rm s}$. These symmetries are spontaneously broken to their diagonal subgroup, $L_{\rm c} \times L_{\rm s} \rightarrow L$, when the gravitational tetrad  emerges as the order parameter of the phase transition.

In this paper we consider the extension of the $PT$ symmetry to thermodynamics, where the $PT$ operation also changes the sign of temperature. We apply this extension to the Dirac vacuum and also to the black holes.

The Universe with negative temperature is obtained using the Dirac picture of the quantum vacuum. The conventional  Dirac vacuum represents an infinite sea of particles with negative energy. Let us note that  that the infinite energy of the Dirac vacuum does not necessarily produce the huge contribution to the cosmological constant. For example, in the so-called $q$-theory, the relevant vacuum energy, which enters the Einstein equations in the form of the  cosmological constant, is determined by the infrared thermodynamics,\cite{KlinkhamerVolovik2008a,KlinkhamerVolovik2008} rather than by the ultraviolet cut-off.  In the fully equilibrium Minkowski vacuum without matter and without external environment the thermodynamic vacuum energy is nullified without any fine tuning. This follows from the thermodynamic Gibbs-Duhem relation, which is valid for any quantum vacuum state, including the non-relativistic vacua -- the ground states in condensed matter. That is why there is no need for the spurious renormalization of the infinite mass of the filled states to zero.

The Dirac picture of the quantum vacuum is also supported by the condensed matter analogy, where the topologically stable  Weyl points in the fermionic spectrum give rise to the massless Dirac spectrum.\cite{Froggatt1991,Volovik2003,Horava2005} In topological superconductors, due to the chiral symmetry the position of the Weyl points is at the zero energy level, giving rise to the Dirac sea of the negative energy states.

\section{Negative temperature in the mirror Dirac vacuum}

If the Dirac picture is adopted, let us apply the parity and time reversal transformation $PT$ to the Dirac vacuum. Here we ignore the breaking of the $P$ and $T$ symmetries in Standard Model, assuming that they are restored at high energy. Then the $PT$ transformation leads to the mirror  Dirac vacuum, where all the positive energy states are occupied and the negative energy states are empty.\cite{Vergeles2021}  The thermodynamic energy of this vacuum remains zero, but the thermal states in the background of this "false" vacuum with inverse population are characterized by negative temperature. 

This is similar to what happens in the subsystem of nuclear spins in condensed matter with inverse spin population.\cite{Lounasmaa1997,Baldovin2021,Braun2013} 
In this spin subsystem  there is an upper limit to all allowed energy states. On the contrary, in the case of Dirac vacua, the energy is unbounded both from below and from above. The origin of the negative temperature and the thermodynamic stability at negative temperature in this case is the exact symmetry between the positive and negative branches of spectrum. 

 At first glance the state with the negative temperature is unstable. However, it is unstable only in case if there is the normal environment -- the thermal bath with positive temperature.  If there is no external environment, i.e. this mirror vacuum occupies the whole universe, this isolated vacuum becomes stable in spite of its negative temperature.

In the relativistic physics, the energy is unbounded. The negative and positive energy branches of fermionic states are  symmetric with respect to zero energy. Due to this symmetry the isolated mirror relativistic vacuum, in which all the positive energy states are occupied and the negative energy states are empty, has exactly the same physics as the conventional Dirac vacuum in the "normal" Universe. 
Though the matter (excitations)  in this mirror Universe has negative energy, and the thermodynamic states are characterized by the negative temperature, the inhabitants of the  mirror Universe would think that they live in the normal Universe with positive energies for matter and positive temperature. It is only with respect to our Universe their temperature and energies are negative. But with respect to their Universe it is our Universe, which is "false" and which is described by negative temperature. 

In principle, we cannot resolve, whether our Universe is "false" or "true". 
Recently there was suggestion to extend our Universe beyond the Big Bang (BB) using the analytic continuation of the radiation-dominated epoch across the BB singularity. \cite{Boyle2018,Boyle2018b} The scale factor $a(\tau) \propto \tau$ changes sign at $\tau=0$. The analytically continued gravitational tetrads also change sign across the BB giving rise to the antispacetime. The same analytic continuation, but which also takes into account the thermodynamics of the matter field, suggests that the temperature on the two sides of the Big Bang has different sign.\cite{VolovikNegT2019} This means that from the point of view of the pre-BB Universe, our post-BB Universe has negative temperature.

So, when one considers the symmetry transformations, one must take into account not only the vacuum states, but also the thermodynamics of the excitations in the background of the vacuum. In particular,  the $PT$ transformation should be accompanied by the reversal of temperature, $T \rightarrow -T$, i.e. $(PT)T=-T$. Here we assume that bosonic fields are composite, being formed as the bi-linear combinations of the fermionic fields, and thus the thermodynamics of bosons follows the thermodynamics of fermions: they inherit  the negative temperature from the fermions.

 It is important that in both Universes, "true" and "false", in spite of their opposite temperatures, the entropy is the same: it is positive. This is because both the energy and temperature of matter change sign, i.e. $\beta(E) = -\beta(-E)$, and thus $S(E) = S(-E)$. 
 For example, for massless fermions, the thermodynamic energy density and the entropy density in two Universes with opposite temperature are:
\begin{equation}
\epsilon = \frac{7\pi^2}{120}  {\rm sign}(T) T^4\,\,,\,\, s =\frac{7\pi^2}{90} |T|^3 \,.
\label{ES}
\end{equation}
 
 \section{Black and white holes in original and mirror vacua}
 
In principle, there can be another situation: when both the temperature and entropy are negative, while energy remains positive.
This situation takes place for thermodynamics of the white holes. The negative entropy of the white hole can be obtained in three different ways of calculations.\cite{Volovik2020a,Volovik2021} In all  three cases  the process of the quantum tunneling from the black hole to the white hole with the same energy (mass $M$) was exploited. Considering the quantum tunneling as thermodynamic fluctuation, and expressing it in terms of the total change of the entropy in this process, one finds the entropy of the white hole. It is with minus sign the entropy of the black hole with the same mass,  and thus the temperature of the white hole is with minus sign the Hawking temperature:
\begin{equation}
S_\text{WH}(M)=- S_\text{BH}(M)\,\,,\,\, T_\text{WH}(M)=- T_\text{BH}(M) \,.
\label{ST}
\end{equation}

In this case the thermodynamic antisymmetry includes also the inversion of entropy, while the energy (mass) is not reversed.
This is because we consider the black and white holes in the same Universe. In the mirror Universe, the black hole partner is also the black hole. This mirror black hole has negative mass, the negative Hawking temperature,  but the positive entropy:
\begin{equation}
T_\text{BH}(-M)=- T_\text{BH}(M) \,\,,\,\, S_\text{BH}(-M)=S_\text{BH}(M) \,.
\label{ST}
\end{equation}
The same is with the white hole. Its mirror partner in the mirror Universe is the white hole with opposite temperature:
\begin{eqnarray}
T_\text{WH}(-M)=- T_\text{WH}(M)=T_\text{BH}(M) \,,
\label{STM1}
\\
S_\text{WH}(-M)=S_\text{WH}(M)=- S_\text{BH}(M)
\,.
\label{STM2}
\end{eqnarray}

\section{Conclusion}

Situation with the negative temperatures in Dirac vacua is actually very similar to the negative temperature in the subsystem of nuclear spins in the case when the nuclear subsystem has the reflection symmetry with respect to the zero energy level. In this case the thermodynamics at $T<0$ will be (anti)symmetric to the thermodynamics at $T>0$. In real nuclear systems such symmetry is absent. For example, the temperature of the magnetic transition at negative temperature does not coincide with the corresponding positive value of the phase transition temperature, $T_{c-} \neq - T_{c+}$, see Ref.\cite{Lounasmaa1997}. 

On the contrary, in the case of Dirac vacua, there is the $PT$ symmetry, which provides the symmetry between the positive and negative branches of the spectrum. That is why, the $PT$ symmetry operations can be extended to include the reversal of temperature. The $PT$ operation  applied to the Dirac quantum vacuum leads to the mirror Dirac vacuum, where all the positive energy states are occupied and the negative energy states are empty. Such vacuum is thermodynamically stable, while the  matter in the background of this vacuum is described by negative temperature.  The thermodynamics in the conventional Dirac vacuum and that in the mirror Dirac vacuum are equivalent, being (anti)symmetric with respect to the reversal of temperature. In particular, the $PT$  symmetry also connects the black hole thermodynamics with that in the vacuum with negative temperature.

All that demonstrates that the equilibrium thermodynamics with negative temperature discussed in Ref.\cite{Baldovin2021} can be realized in quantum field theory.

  {\bf Acknowledgements}. I thank M. Baldovin and S.N. Vergeles for discussions. This work has been supported by the European Research Council (ERC) under the European Union's Horizon 2020 research and innovation programme (Grant Agreement No. 694248).

\end{document}